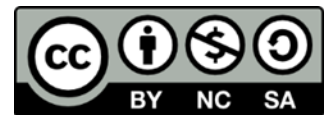

# Choreography in the Embedded Systems Domain: A Systematic Literature Review

Nebojša Taušan[1], Jouni Markkula, Pasi Kuvaja, Markku Oivo

University of Oulu, M3S, P.O. Box 3000, Oulu, Finland

Abstract

**[Context]** Software companies that develop their products on a basis of service-oriented architecture can expect various improvements as a result of choreography. Current choreography practices, however, are not yet used extensively in the embedded systems domain even though service-oriented architecture is increasingly used in this domain. **[Objective]** The objective of this study is to identify current features of the use of choreography in the embedded systems domain for practitioners and researchers by systematically analysing current developments in the scientific literature, strategies for choreography adaption, choreography specification and execution types, and implicit assumptions about choreography. **[Method]** To fulfil this objective, a systematic literature review of scientific publications that focus on the use of choreography in the embedded systems domain was carried out. After a systematic screening of 6823 publications, 48 were selected as primary studies and analysed using thematic synthesis. **[Results]** The main results of the study showed that there are differences in how choreography is used in embedded and non-embedded systems domain. In the embedded systems domain, it is used to capture the service interactions of a single organisation, while, for example, in the enterprise systems domain it captures the service interactions among multiple organisations. Additionally, the results indicate that the use of choreography can lead to improvements in system performance and that the languages that are used for choreography modelling in the embedded systems domain are insufficiently expressive to capture the complexities that are typical in this domain. **[Conclusion]** The selection of the key information resources and the identified gaps in the existing literature offer researchers a foundation for further investigations and contribute to the advancement of the use of choreography in the embedded systems domain. The study results facilitate the work of practitioners by allowing them to make informed decisions about the applicability of choreography in their organisations.

---

[1] Corresponding author: Nebojša Taušan; Mobile: +381 63 557 365; Postal address: P.O. Box: 3000, 90014; University of Oulu; Finland; E-mail: nebojsa.tausan@oulu.fi

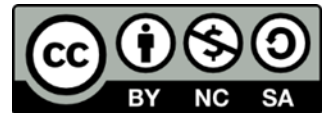

# 1 Introduction

The growing utilisation of embedded systems (ES) in different application areas, such as automotive, avionics and home appliances, has led to the growth of their software size and complexity [1]. As a consequence, ES have become inflexible, thus difficult to maintain, while the associated development costs and time have increased. One approach to addressing these challenges has been the adoption of service-oriented architecture (SOA), which originates from the enterprise systems domain and has been designed to tackle the corresponding challenges. SOA has been successfully applied in enterprise systems development to advance the flexibility and to reduce the costs and time required for development [2][3][4][5]. Consequently, SOA is increasingly adopted in the ES domain, [6][7][8][9].

Enterprise systems built following SOA can be seen as a collection of software services built by independent organisations or companies that have common business goals [10]. These systems support the realisation of business goals through the service interactions. Since doing so represents the realisation of the organisations' goals, the specification of service interactions has to include sufficient details for its users, and thus it is often specified from different viewpoints [11].

Different viewpoints focus on different sets of service interaction details and use their own set of concepts to document those details. Choreography represents a viewpoint on service interactions in SOA-based development [12]. In the enterprise domain, choreography specifies the sequence of interactions that realises the organisations' business goals. In an ES domain, choreography is used in the same manner, but instead of focusing on the organisations, it specifies the sequence of interactions among devices in a product of which they are a part. For example, a car includes a number of devices that are responsible for tasks such as fuel control and the braking system. These devices are comparable to organisations in the enterprise domain and thus seen as service providers whose services are interacting to fulfil the goals such as automatic cruise control.

A number of studies focusing on choreography in ES have been published. For example, the documentation of the smart embedded devices in manufacturing is expected to include a choreography specification [13]. Specifying choreography can also contribute to the agility, and thus to the flexibility, of industry supply chain systems [14]. The analysis of the acceptance and use of choreography among practitioners indicated their interest in using choreography and that choreography tools can increase their productivity [15]. The results of these studies indicate the growing importance of choreography in ES, however, the results are also scattered and there is neither a common understanding of how to use the choreography in ES nor a consensus on how it affects the ES development process. For this reason, there is a need to gather the existing information in the scientific literature, to analyse it systematically and present a synthesis of the current knowledge of choreography usage in the ES domain. This need motivated the objective of this study and the selection of the research method.

The objective of this study is to identify the current features of the use of choreography in the ES domain that are relevant for practitioners and researchers. To accomplish this, the existing scientific literature was systematically analysed considering current developments, strategies for choreography adaption, choreography specification and execution types, and implicit assumptions about choreography. The study was carried out by applying a systematic literature review [16], which is an accepted method for synthesising knowledge from the existing scientific literature.

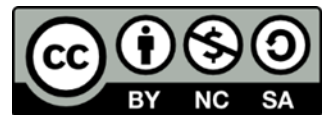

This article is structured into six sections. Section 2 presents the study background by describing the concepts relevant to this study. Section 3 describes the literature review protocol that was followed during the collection of relevant studies and how the studies were analysed. The analysis results are presented in Section 4, and this presentation is divided into five subsections, each presenting an answer to one of the research questions. Section 5 presents the study findings, while Section 6 presents the validity threats. The study conclusions are presented in Section 7.

## 2   Background

*SOA* is an approach to structure or organise software systems [17][10]. The main building block of SOA-based systems is the *service,* which represents a software implementation of a clearly defined task. Services are provided by participants, the meaning of which depends on the domain where they are used. In an enterprise domain, participants represent an individual organisation or a company. In an ES domain, it represents a device or a component within a larger product. Participants have common goals and collaborate to fulfil these goals. SOA-based systems support this collaboration through service interactions. Due to the relation between the fulfilment of participants' goals and service interactions, the specification of service interactions is seen as an important part of the overall documentation of SOA-based systems. Two viewpoints should be considered to specify the interactions – choreography and orchestration [11][12].

*Choreography* focuses on the flow of interactions between different participants that are involved in collaboration and presents the interactions from a global or neutral viewpoint. *Orchestration* shows service interactions from the viewpoint of a single participant involved in collaboration. These viewpoints capture the interactions from different abstraction levels [11]. The choreography represents the highest abstraction level since it is a global viewpoint on the sequence of interactions [18]. Once specified, choreography provides details about: the realisation of participants' common goals; a contractual agreement between participants on their common behaviour and a communication tool for agreeing on different development-related questions. To specify the interactions from the global point of view, choreography uses constructs, the most important of which are as follows:

- Participant: Service provider involved in collaboration to fulfil its goal.
- Role: More closely describes the work of a participant within the concrete choreography specification and enables better comprehensiveness due to richer semantics.
- Interaction: The main construct in a choreography specification. Each interaction in a sequence captured by a choreography specification represents an exchange of information between two or more participants.
- Control structures: Denote the rules that govern the flow of interaction between the participants. These structures include interaction sequencing, (multiple) branching, parallel execution and iterations.

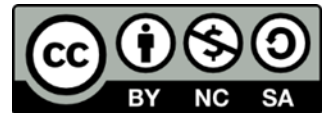

Beside these constructs, a choreography specification can contain information about other development artefacts, such as the messages that is exchanged between the participants, the events that can occur during interactions and the message exchange patterns [19].

Interactions can also be specified from the orchestration point of view. Orchestration specifies interactions from the perspective of an individual participant and captures the service interactions needed to achieve a single participant's goal. This differs from choreography, which captures the interactions between participants that realise their common goal and makes orchestration a local view or a projection on choreography from the perspective of a single participant that is engaged in that choreography. Choreography and orchestration therefore represent two complementary viewpoints that specify the interactions from different abstraction levels.

Services and interactions, as the fundamental concepts in SOA, are increasingly adopted in the ES domain [6][7][8][9]. *ES* is seen a software-controlled device designed with a dedicated function in mind, but it can also be seen as a function or service provider within a larger product such as automobiles, trains or aeroplanes [20]. Consequently, these products can combine services that are provided by distinct ES according to the goal they aim to fulfil. Our perspective is that with the increase in the number of networked ES, or participants, the need to specify their interaction from a global or choreography viewpoint will increase accordingly, while the need for orchestration will remain unchanged and limited to devices and networks with excessive resources and processing capabilities. Besides the scientific literature, these reasons additionally contribute to our decision to focus on choreography and how it is utilised in the ES domain.

# 3 Research Method

This study followed the systematic literature review (SLR) guidelines proposed by Kitchenham and Charters [16] consisting of three major phases – planning, conducting and analysis. The research questions and the major decisions about the search protocol were made during the planning phase. The search, selection and quality assessment of the primary studies were completed during the conduction phase, while the study results were derived in the analysis phase.

## 3.1 Planning Phase

The planning phase of the study consisted of the following decisions: specification of research questions, the selection of publication sources, keywords and criteria for publication inclusion or exclusion, the piloting of keywords and criteria and the definition of quality assessment procedures.

**The research questions** (RQs) that guided this study are as follows:

- RQ1: What are the current developments in the scientific literature that focus on choreography in ES?
- RQ2: How is choreography adapted in the ES domain?
- RQ3: How is a choreography specification used in the ES domain?
- RQ4: How is a choreography specification realised during the ES execution?

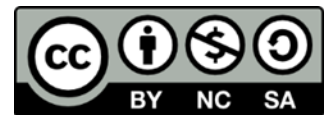

- RQ5: What implicit assumptions regarding choreography should be considered when utilising choreography in the ES domain?

RQ1 contributes to the study's objective by providing a detailed overview or a map of the studies that focus on choreography in the ES domain. RQ2 contributes to the study's objective by exploring various choreography adaption strategies in the ES domain. RQ3 and RQ4 contribute to the study's goal by presenting how a choreography specification is used in the ES development process and how it is realised during ES execution. RQ5 identifies implicit assumptions that should be considered when utilising choreography in the ES development domain since they can influence various design decisions.

**Literature sources and keywords.** Broad coverage of the literature published in the software engineering field was ensured by using the following sources: ACM Digital Library, CiteSeerX, Google Scholar, IEEE eXplore Digital Library, ProQuest, World of Science, Scopus and Springer. The keywords and the search string used for querying the publication sources are presented in Table 1. These keywords were selected based on terminology that is used in the scientific literature that was known to the authors before the present study and the analysis of two relevant embedded systems web sites (machinetomachinemagazine.com, embedded.com).

| Keywords and search string | |
|---|---|
| Part I | choreography AND |
| Part II | embedded OR automotive OR telecom* OR automation OR healthcare OR aerospace OR robotics OR 'internet of things' OR 'web of objects' OR mobile |
| Part III | AND service-oriented OR SOA |

**Table 1 Keywords and the search string**

The keywords in parts I and III of the search string aimed to constrain the search only to publications that include choreography and SOA-based systems. The keywords in part II aimed to constrain the search to studies focusing on the ES domain.

**Criteria for publication inclusion or exclusion.** Three groups of criteria were derived for including and excluding publications. The first group of criteria is based on the bibliographic data. Based on this group, a publication was excluded in cases when 1) the written language was not English, 2) it was a duplicate entry, 3) the data were erroneous, 4) it was non-peer reviewed, 5) it was a textbook, 6) it was an MSc or PhD theses, 7) it was a proceedings preface, keynote or panel discussion or 8) it was not from the software engineering field.

The second group of criteria is based on the abstracts of the publications. Predefined sets of keywords, organised in four groups, and presented in Table 2, were used. If at least two keywords belonging to the choreography, SOA and ES groups were found, that publication was accepted in the list. If at least one keyword belonging to the enterprise systems group was found, the publication was excluded from the list. During the evaluation of abstracts, the researchers used a predefined evaluation scale consisting of three values — 'accept', 'exclude' and 'cannot decide'.

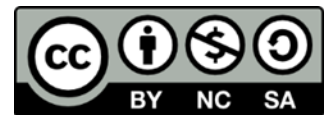

| Include if: | | | Exclude if: |
|---|---|---|---|
| Choreography | SOA | Embedded systems | Enterprise systems |
| Conversation, Composition, Orchestration, Interaction, Process collaboration | SOA, Service, Service-orientation, Service-based | Telecom, Automotive, Automation, Manufacturing, Robotic, Mobile, Smart, Internet-of-things, Web-of-objects, New-generation network NGN, Sensor networks, Real-time systems, Ubiquitous, Pervasive, Traffic control systems, Sensor networks, Process control | Information system, Business system, Enterprise-resource planning or ERP, Customer-relationship management or CRM, Supply-chain management or SCM, Decision-support system, Expert system, Enterprise integration application or EIA, E-government, e-commerce, web application |

Table 2 Keywords for the inclusion and exclusion of publications based on their abstracts

The third criterion represents a simple heuristic. According to this heuristic, researchers were to check the illustrations (examples, schemas, graphs, diagrams) presented in the study. If the main participants in the illustrations come from the enterprise domain (e.g. buyer, seller, supplier, bank, insurance organisation), that publication was excluded from the list. Similarly, if the main participants originate from the ES domain (e.g. sensors, actuators or robots) the publication was accepted in the list.

**Piloting of keywords and criteria.** Before the publication search, the keywords and criteria were piloted to ensure their suitability for this study. To pilot the criteria for abstract-based acceptance and exclusion, the researchers first randomly selected 60 publications and then evaluated those independently using the defined criteria. The comparison of evaluation results revealed inconsistencies, the cause of which is traced to the terminology used in the criteria. Following this, the criteria were updated and final guidelines for the evaluation were compiled. In the second round of piloting, the researchers evaluated 60 publications equally in 90% of the cases.

**Quality assessment procedure.** To assess the quality of the selected primary studies, the criteria used in the SLR study by Nicolas and Toval [21] were adopted. These criteria include the assessment of the following:

- Publication place. Studies must originate from sources that publish scientific literature in the software engineering field.
- Artefact in focus. Studies must focus on the choreography or on the artefact that is clearly related to the choreography.
- Validation procedures. The validation of the study results must fit into one of the following levels: a) the study is presented in the academic context or using examples, b) the study is introduced to practice as part of an industrial case study, c) the study can be seen as part of industrial practice and d) the study is validated using known research methods.

In this study, the quality assessment deviated from the criteria proposed by Kitchenham et al. [16] for two reasons. First, the use of choreography in the ES domain is still a developing topic, meaning that there is a lack of consensus on how to study the phenomenon. A thorough assessment of the applied research methods would lead to the exclusion of a number of relevant studies and cause the loss of the needed information. Second, the studies are strongly design- or engineering-oriented and focused on solving practical problems by designing artefacts. These studies tend to put more emphasis on the methodological validation of their artefacts and less on describing the methods that guided their design process. Since the validation procedures are among the quality criteria used in Nicolas and Toval SLR [21], the adoption of these criteria in this review was seen as a reasonable choice.

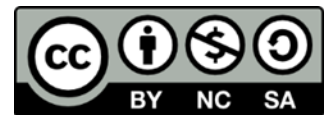

## 3.2 Conducting Phase

During the conducting phase, literature sources were searched, screened and selected following the protocol presented in Figure 1. The protocol steps are presented as rectangles, the criteria as dotted-line ovals while the full-lined ovals represent the publication list.

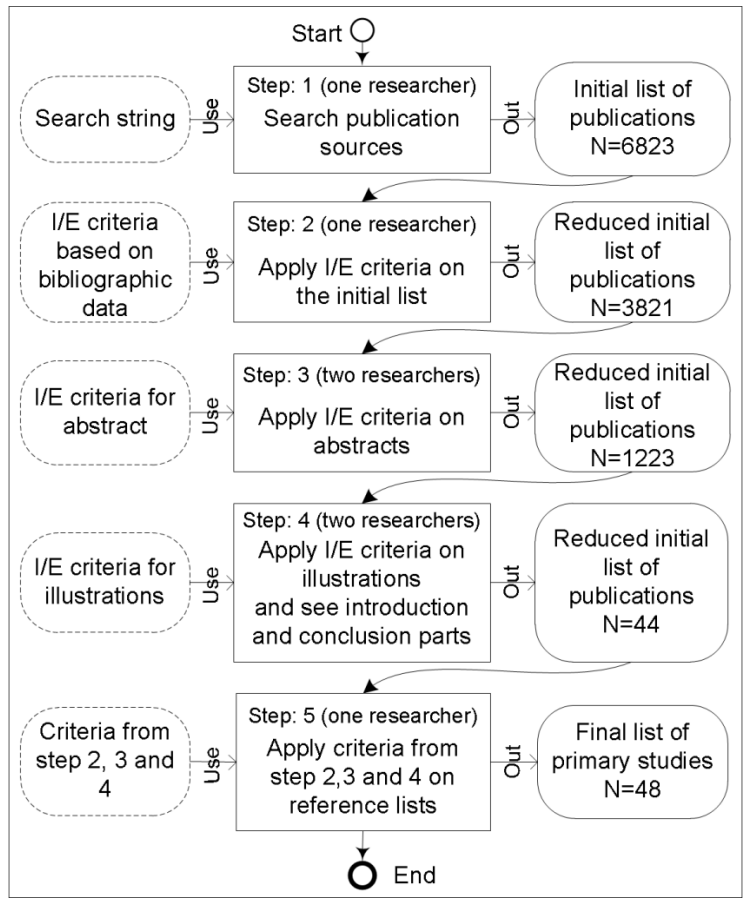

**Figure 1 Search protocol**

The search protocol involved the following steps:

**Step 1.** A single researcher queried the defined sources using the defined search string as the criteria. The search execution date was September 2015, and this search retrieved the bibliographic data of 6823 publications.

**Step 2.** The data from the initial list were checked against the inclusion/exclusion criteria, by a single researcher, based on the bibliographic data. This excluded 3002 publications, reducing the list to 3821 titles.

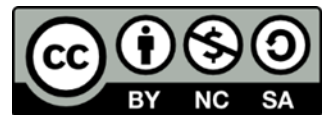

**Step 3.** Two researchers independently read the abstracts and assigned them a predefined evaluation mark. According to an agreed-upon rule of accepting or excluding a publication, the decision was made if both researchers 'accepted' or 'excluded' the publication, or if one of the researchers 'could not decide' while the other accepted or excluded it. In cases when both researchers could not decide or when one researcher decided to accept while the other to exclude, the publication was accepted in the list. After completing this step, 2599 publications were excluded, leaving 1223 publications on the list.

**Step 4.** The introductions, conclusions and illustrations of the remaining publications were read. A simple heuristic, defined for this step, was used for guiding the acceptance/inclusion decision. This step was conducted by two researchers; the first researcher read these parts of each publication and decided to accept or to exclude the publication, while the second researcher randomly selected a set of publications and verified the decision. As a result, a list was formed containing 44 primary studies.

**Step 5.** Scientific publications that were cited in the 44 publications that remained on the list were screened. During this step, a single researcher successively applied the criteria used in steps 2, 3 and 4 to the cited references. This resulted in four publications being added to the list. This step concluded the search protocol and resulted in a primary study list containing 48 scientific publications.

The quality of the selected primary studies was assessed using the specified quality criteria. This assessment resulted in acceptance of all the 48 studies as suitable for the analysis. The studies originate from a credible publication source, have choreography as their focus and are verified in an academic environment using examples (32 studies, 67%), using research methods (8 studies, 17%), the industrial case (6 studies, 12%) and as an industrial practice (2 studies, 4%). Detailed results of the quality assessment are presented in Appendix I.

## 3.3 Analysis Phase

The thematic synthesis approach [22] and the analytical tool developed by Leite et al. [23] were used for the analysis of the studies. Thematic synthesis relies on techniques such as coding, category and theme development for the analysis and synthesis of new knowledge. The adopted analytical tool by Leite et al. [23] is specifically developed to characterise the choreography adaption.

### 3.3.1 The thematic synthesis

The thematic synthesis approach consists of five main steps. The first step is data extraction, during which all data relevant to answering the RQs are extracted. The second step is coding, which denotes the linking of the labels, also known as codes, with the pieces of text seen as relevant or logically related to a specific code. Theme derivation is the third step and denotes the translation of codes to higher-order themes. The fourth step is the model derivation, which denotes the development of the conceptual model that summarises the themes and categories derived in the previous step. The fifth step denotes the assessment of the results' trustworthiness. During the execution of these steps, the researchers used NVivo software [24], which supported the analysis tasks, such as coding, category development, content organisation, complex searches, topics linking and content annotation.

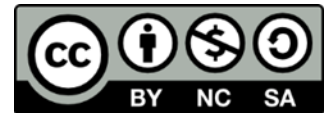

**Data extraction.** The data extracted to answer the RQs are presented in Table 3. For RQ1, the data are divided into two groups: a bibliography that characterises the publications and research that characterises the study reported in the publications. For RQ2, the data are also divided into two groups: strategies and aspects. The data for answering RQ3, RQ4 and RQ5 consist of a single data point per RQ.

| Data group | Data point | Target RQ |
|---|---|---|
| Bibliography | title, authors, year, publication channel | RQ1 |
| Research | research project, industry branch, context, motivation, modelling languages, development environment, study impact, validation procedures, is-quantitative, future trends | |
| Strategies | model-driven engineering, measurement, multi-agent, formal, semantic-based, middleware | RQ2 |
| Aspects | target, intervention degree, scalability, implementation, models | |
| | use of choreography specifications | RQ3 |
| | Realisation of choreography specifications | RQ4 |
| | Implicit assumptions for choreography use | RQ5 |

**Table 3 Data extraction form**

**Coding.** The researchers applied integrated coding, combining deductive, or top-down, and inductive, or bottom-up, development of codes [22]. Integrated coding starts with an initial list of codes, which evolves during the coding process. The data points used to answer RQ1 and RQ2 were encoded in a deductive or top-down manner. The data points for RQ3–RQ5 were encoded in an inductive or bottom-up manner. Inductive coding evolved the initial list, presented in Table 3, and the results are presented in Table 4.

| Code group (category) | Code | Target RQ |
|---|---|---|
| Utilisation types of choreography specifications in ES | descriptive, parsed, executable | RQ3 |
| Run-time realisation of choreography specifications in ES | orchestrated, distributed, piggybacked | RQ4 |
| Participant autonomy | autonomous, non-autonomous | RQ5 |
| Network stability | stable, ad-hoc | |

**Table 4 Changes to the initial list of codes**

**Theme derivation.** Based on the analysis, five higher-order themes were derived. Each one is related to a corresponding RQ. The derived themes were: current developments; choreography adaption; the utilisation types of choreography specifications in ES; the run-time realisation of choreography specifications in ES and implicit assumptions for using choreography in ES. To derive the choreography adaption theme, during this step, the analytical tool was populated with the extracted data (strategies and aspects data groups in Table 3) and analysed.

**Model derivation.** The development of a conceptual model was based on the derived themes. It includes themes, their categories and the relevant codes, which altogether summarise the characteristics of the utilisation of choreography in the ES domain. The derived model is presented in Figure 2 where the themes are represented in highlighted ovals and the categories in rectangles.

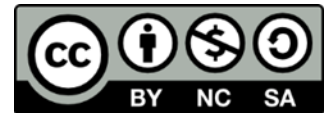

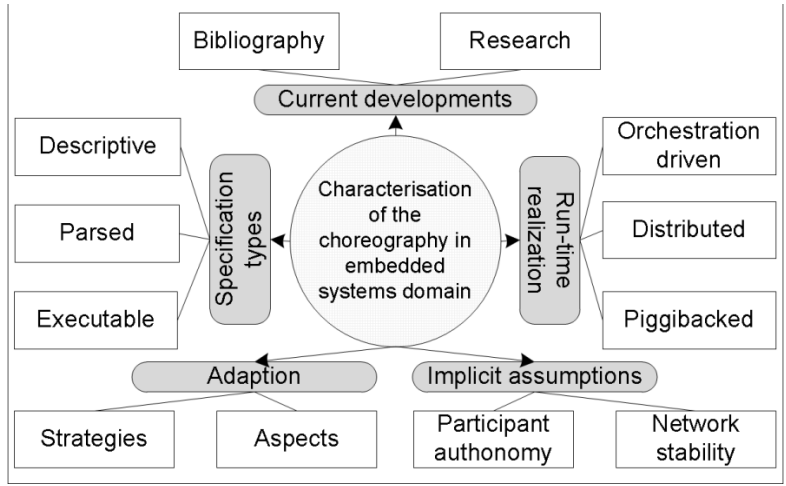

Figure 2 Conceptual model of themes and categories

**Assessment of trustworthiness.** The trustworthiness of the results was ensured by addressing the descriptive, theoretical and construct validity, generalisability and repeatability. A detailed description of how these validity threats were addressed in this study is presented in Section 6.

### 3.3.2   Choreography Adaption Analysis Tool

For the analysis of the choreography adaption, to answer RQ2, an analytical tool developed by Leite et al. [23] was used. The analytical tool represents a matrix that arranges six types of adoption strategies and five strategy aspects that further refine each of the adaption strategy types. Table 5 describes these strategies and aspects.

| Adaption strategy types | Aspects of the adaption strategy |
|---|---|
| I. Model-driven engineering-based: Relies on models, model transformation and automatic code generation to perform the adaption. | 1. Target: Indicates whether the strategy support changes in functional or non-functional requirements. |
| II. Measurement-based: Adaption is automatically triggered when a given threshold is violated by the choreography. | 2. Required intervention degree: Indicates if the adaptation is performed automatically or by human intervention. |
| III. Multi-agent-based: Adaption is enabled using the agents that are capable of learning, analysing and autonomous decision making. | 3. Scalability impact: Indicates if the scalability of the choreography specification is discussed in the context of strategy. |
| IV. Formal method-based: Adaption that relies on process calculi or finite state machines. | 4. Implementation: Indicates whether the strategy is implemented by a tool or a prototype. |
| V. Semantic reasoning-based: Adaption uses the ontologies to reason about the communication among the services. | 5. Models: Indicates which choreography models, representations or standards are used in the strategy. |
| VI. Middleware-based: Adaption uses the middleware features to intercept service messages and to make the necessary adoptions. | |

Table 5 Analytical tool for the analysis of the choreography adaption strategies (source: [23])

The analysis revealed which adaption strategies are most commonly used in the ES domain, what aspects they encompass and what the differences are between the results in SLRs. Based on the analysis results, RQ2 was answered.

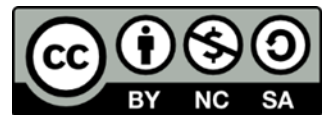

## 4   Results

Five themes that represent the answers to the stated RQ were derived based on the analysis of the identified studies. The first theme presents current developments in studies focusing on choreography in ES. The second theme presents choreography adaption in the ES domain, while the third and fourth reveal how a choreography specification is used during the ES development and how it is realised during ES execution. The fifth theme describes implicit assumptions that should be considered when using choreography in ES. These themes were derived based on the analysis of the studies focusing on choreography in ES. The reference numbers that identify each study, the authors and the publication year are presented in Table 6.

| Reference | Authors | Year | Reference | Authors | Year |
|---|---|---|---|---|---|
| [25] | de Souza, Victor A S M and Cardozo, Eleri | 2005 | [49] | Dar, Kashif et al. | 2011 |
| [26] | Sadok, Eamil F and Liscano, Ramiro | 2005 | [50] | Zayati, Ahlem et al. | 2012 |
| [27] | Cottenier, Thomas and Elrad, Tzilla | 2005 | [51] | Minor, Johannes et al. | 2012 |
| [28] | Cambronero, María-Emilia et al. | 2006 | [52] | Cherrier, Sylvain et al. | 2012 |
| [29] | Ferrari, Gianluigi et al. | 2006 | [53] | Starke, Günther et al. | 2013 |
| [30] | Delamer, Ivan M and Lastra, Jose L Martinez | 2006 | [54] | Cherrier, Sylvain et al. | 2013 |
| [31] | Delamer, Ivan M and Lastra, Jose L Martinez | 2006 | [55] | Cortes-Cornax, Mario et al. | 2013 |
| [32] | de Souza, Victor A S M and Cardozo, Eleri | 2006 | [56] | Hu, Raymond et al. | 2013 |
| [33] | Rossebø, Judith E Y and Runde, Kobro | 2008 | [57] | Kattepur, Ajay et al. | 2013 |
| [34] | Sen, Rohan et al. | 2008 | [58] | Herrera, Edwin Cedeno et al. | 2013 |
| [35] | Chan, Pat and Lyu, Michael R | 2008 | [59] | Taušan, Nebojša et al. | 2013 |
| [36] | Phaithoonbuathong, Punnuluk et al. | 2008 | [60] | Ramis, Borja et al. | 2013 |
| [37] | Zhang, Zhuoyao et al. | 2008 | [61] | March, Verdi et al. | 2013 |
| [38] | Rossebø, J E Y and Bræk, Rolv | 2008 | [62] | Cherrier, Sylvain et al. | 2014 |
| [39] | Kirkham, Thomas et al. | 2008 | [63] | Tanveer, Ahmed & Abhishek, Srivastava | 2014 |
| [40] | Jie, Guo et al. | 2009 | [64] | Hu, Jianpeng et al. | 2014 |
| [41] | Pedraza, Gabriel and Estublier, Jacky | 2009 | [65] | Cherrier, Sylvain et al. | 2014 |
| [42] | Fernández-Llatas, Carlos et al. | 2010 | [66] | Kothmayr, Thomas et al. | 2014 |
| [43] | Mersch, Henning et al. | 2010 | [67] | Kaur, Navjot et al. | 2015 |
| [44] | Mostarda, Leonardo et al. | 2010 | [68] | Duhart, Clément et al. | 2015 |
| [45] | Ciancia, Vincenzo et al. | 2010 | [69] | Cherrier, Sylvain et al. | 2015 |
| [46] | Honda, Kohei et al. | 2011 | [70] | Kothmayr, Thomas et al. | 2015 |
| [47] | Ciancia, Vincenzo et al. | 2011 | [71] | Dar, Kashif et al. | 2015 |
| [48] | Cherrier, Sylvain et al. | 2011 | [72] | Leitão, Paulo et al. | 2015 |

Table 6 Primary studies

In total 48 studies focusing on choreography in ES were selected and analysed in this SLR. The reference numbers that identify each study, the authors and the publication year are presented in Table 6.

### 4.1   Current developments in studies focusing on choreography in ES

To answer RQ1 regarding the current developments in studies focusing on choreography in ES, the bibliographic data and the data that characterise the research reported in these studies were analysed. The data, characteristics and categories were derived during the data collection, coding and theme derivation steps of the thematic synthesis approach used in this SLR to analyse the primary studies.

#### 4.1.1   Characteristics of the bibliographic data

The analysed bibliographic data included the publication sources, publishing channel and the publication year. The frequencies of studies according to these data were calculated and are summarised in Figure 3.

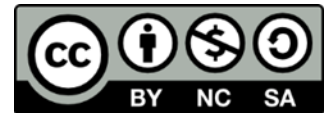

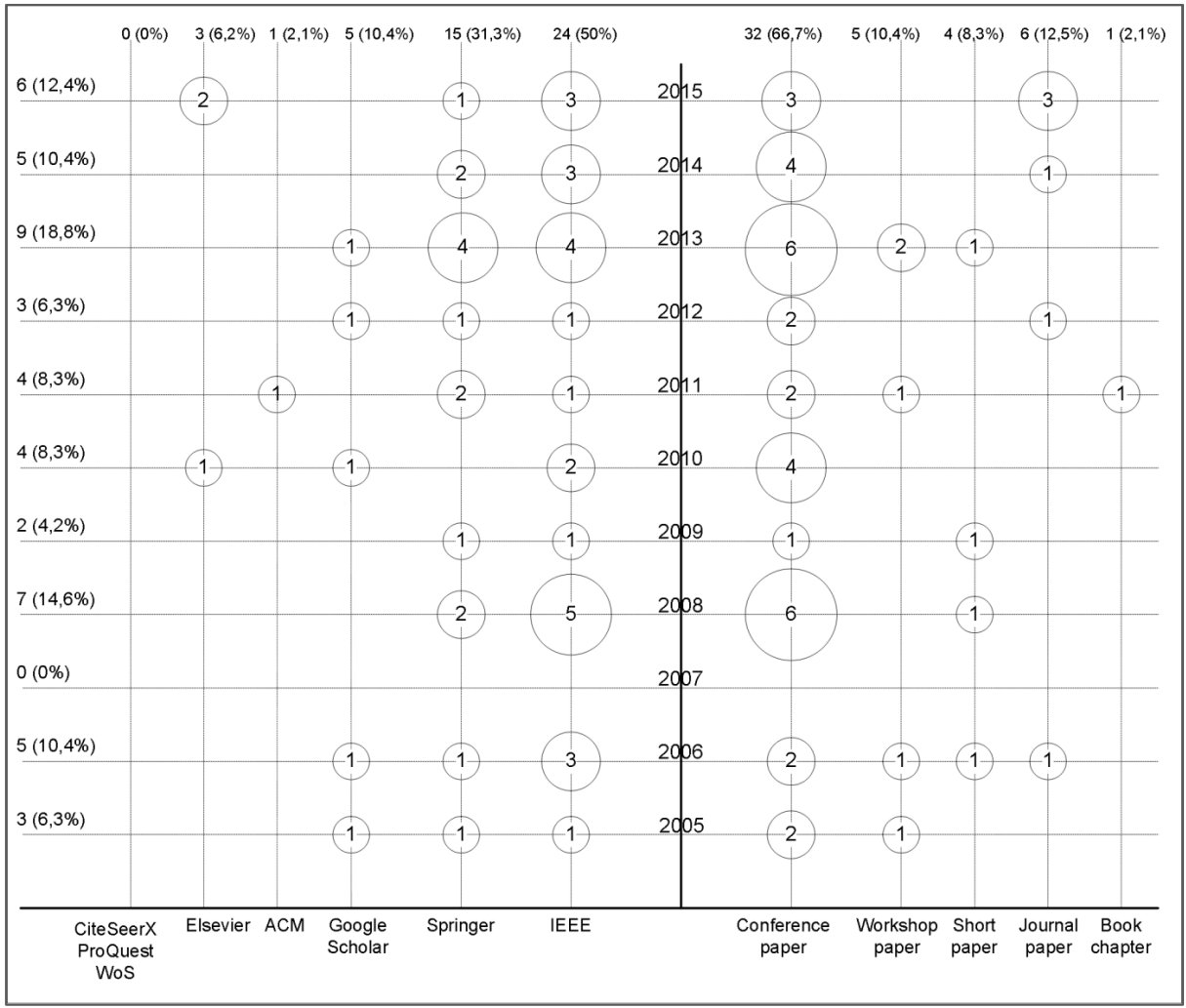

**Figure 3 Publication sources and publishing channels**

Based on the frequencies presented in Figure 3, the following characteristics can be derived:

- Primary studies were published over a period of 10 years (from 2005 to 2015), and the largest number of studies was published in 2013 (9, 18.8%).
- The publication sources from which most of the studies originate are IEEE (24, 50%) and Springer (15, 31.3%).
- Conference papers are the most frequently used publishing channel (32, 66.7%), while scientific journals are moderately used (6, 12.5%).

The relatively low number of journal publications can be explained by acknowledging that choreography in ES is a developing topic. This explanation can be supported with two arguments. The first argument is the 10-year lifetime of this topic, which can be considered insufficient to reach sufficiently matured development and research levels. The second argument is that the terminology used in choreography related publications remains unclear and often used with conflicting meaning. This argument can also be supported with findings from [23].

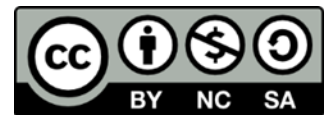

### 4.1.2 Characteristics of the Reported Research

Seven characteristics of the research reported in the studies were analysed. These characteristics and the reasons they were analysed are as follows:

1) Research projects: Reveals the research projects during which the studies focusing on choreography in ES were conducted. The purpose of this analysis is to briefly present the topics of these projects as wider areas where the choreography is studied.

2) Industry branch: Reveals the industry branches where the choreography is studied during ES development. The purpose of this analysis is to present the industry branches that are showing interest or already use the choreography in their work.

3) Choreography context: Reveals the software development lifecycle artefacts, such as methods, ontologies or middleware, alongside which the choreography is studied. The purpose of this analysis is to identify the lifecycle artefacts were the choreography is studies and alongside which artefacts the choreography is studied the most.

4) The study's motivation: Focuses on the ES development problems that motivated the research reported in the studies. The purpose of this analysis is to generalize the types of problems in ES development that can be addressed with choreography.

5) Choreography tools: Choreography tools in this analysis denote languages used to specify choreography and software packages that support the choreography work. The purpose of this analysis is to provide insight into which tools are most commonly used in the studies.

6) The study's impact: Investigates how the results of each study impact the ES development process, quality attributes of a software product and the product's performance. The purpose of this analysis is to relate the use of choreography with and the resulting benefits.

7) Future trends: Reveals what is reported as being the next steps for future research. The purpose of analysing this characteristic is to provide a prediction of the future advancement of this topic.

The analysis of the studies according to the above characteristics resulted in a categorisation scheme for each of the characteristics. These categorisation schemes, study frequencies and the results of the analysis are presented in the following text.

**Research projects.** Ten research projects were found that explored the utilisation of choreography in ES. These research projects were funded by three sources: the European Framework Program[2] (EFP), ITEA[3] and national funding organisations. More details about the projects are summarised in Table 7.

---

[2] cordis.europa.eu/fp7
[3] itea3.org

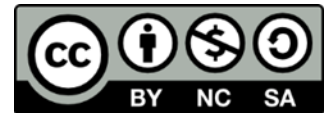

| Funding Program | Project | Project reference | Count ries | Particip ants | Primary study | Description |
|---|---|---|---|---|---|---|
| EFP | PLANTCockpit | [73] | 7 | 13 | [60] | SOA principles and practices for enhancing productivity in industry manufacturing |
| | SOCRADES | [74] | 6 | 16 | [39] [53] | Comprehensive SOA-based approach to integrating smart ES |
| | IMC-AESOP | [75] | 7 | 17 | [53] | SOA approaches for process control applications in large industry systems |
| | CHOReOS | [76] | 6 | 14 | [57] | Framework for scalable choreography in the context of the future internet |
| | SENSORIA | [77] | 7 | 19 | [45] [47] | Foundational theories, techniques and methods for SOA-based systems |
| | VAALID | [78] | 4 | 8 | [42] | Approach to the design of service-based solutions in the context of ambient-assisted living |
| ITEA | AMALTHEA | [79] | 3 | 15 | [59] | Tools and platform for integration in the context of automotive software development |
| | SODA | [80] | 7 | 27 | [39] | Ecosystem to manage interworking and communications between devices in the SOA context |
| National funding | SARDAS | [81] | 1 | 4 | [33] [38] | Methodology for robust design, assessment and specification to ensure availability |
| | IMPRONTA | [82] | 1 | 5 | [30] [31] | Industrial Manufacturing Platform for Reconfigurable, Agent-based Production |

Table 7 International research projects

The analysis of the research projects reveals that, in most cases, choreography was studied alongside the SOA and its application in ES industries. This is understandable since choreography represents a viewpoint on service interactions in SOA-based systems. One exception is the CHOReOS project that focuses specifically on choreography and how it is used in the context of future internet. The work on this particular project is also continued in the Horizon 2020[4] project CHOReVOLUTION [83]. The work on the CHOReVOLUTION project, supported with a significant number of countries and project participants presented in Table 7, provides support for the attractiveness of this topic and serves as a motivation for further studies.

**Industry branch.** The analysis of the studies identified seven industry branches in which choreograph is studied and used for ES development. These industry branches were identified based on their description in the studies, thus it should be noted that two categories (embedded systems and sensors and actuators) differ from others since they encompass studies whose results are not limited to a specific industry branch. The contributions from the studies under these categories can be seen as transferable and considered an addition to studies belonging to other industry branches. The distribution of studies based on the classification is presented in Table 8.

---

[4] http://ec.europa.eu/programmes/horizon2020/en

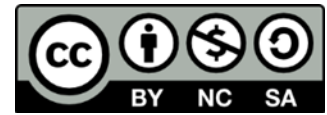

| Industry branch | Frequency (N=48) | Primary studies | Example of ES from industry branches |
|---|---|---|---|
| Industry automation | 14 (29%) | [30][31][36][39][42][43][50][51][53][60][66][67][70][72] | Industry robots, conveyors, lifters and positioning systems |
| Internet of things | 11 (23%) | [34][49][48][54][57][58][62][63][65][69][71] | Various devices connected to the internet, such as smartphones and home entertainment systems |
| Telecom and networking | 9 (19%) | [25][29][32][33][37][38][40][45][61] | Routers, switches, radio antennas |
| Embedded systems | 6 (13%) | [27][28][41][46][56][59] | Computing systems embedded in a larger machine to control one or more of its dedicated functions |
| Sensors and actuators | 4 (8%) | [26][44][52] [68] | Air flow, temperature, pressure meters, injectors, igniters and valve controls |
| Aerospace | 3 (6%) | [35][55] [64] | Navigation, landing and power control systems. |
| Automotive | 1 (2%) | [47] | The vehicle engine, steering, brake and suspension control systems |

Table 8 Primary study distribution according to industry branch

Table 8 shows that different industry branches are using or have interest in using choreography for ES development. The presented study frequencies show that the leading branches studying choreography are industry automation and the Internet of things. Strong interest in the application of SOA principles in industry automation, as evidenced by several research projects (PLANTCockpit, IMC-AESOP, IMPRONTA), is seen as a possible explanation for the large number of studies in this branch. One possible explanation for the growing interest in choreography in the Internet of things domain is its ability to capture the distribution of roles and responsibilities among the autonomous devices, or things, which are interacting over the public or private network. Additionally, devices and networks in the Internet of things domain are often energy restricted, impeding the use of centralised orchestration engines and thus limiting the need for orchestration viewpoint modelling.

**Choreography context.** The analysis of the choreography context in the studies resulted in a categorisation scheme consisting of six categories.

- Method: Choreography is discussed as part of an ES development method. The method in this categorisation encompasses the proposed guidelines, approaches, frameworks, architectures that target different stages and artefacts in ES development.
- Middleware: Studies that discuss choreography in the context of middleware products.
- Modelling Language: Studies that discuss the development, extension, description or other activities related to languages for choreography modelling and execution.
- Tools: Studies that report on the current achievements in the development of software tools that support various aspects of the ES development.
- Ontologies: Studies that propose different ontology models to supplement the semantics of languages for choreography modelling.

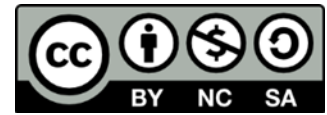

- Comparisons: Studies where two or more alternatives are compared to understand which alternative is more suitable in a given context. Some examples include the comparison of modelling languages, interactions schemes and techniques.

The studies organised according to the choreography context are presented in Table 9. Note that some of the studies are included in more than one category since they investigate choreography alongside more than one lifecycle artefact which is included in the categorisation.

| Context | Frequency | Primary studies |
|---|---|---|
| Method | 24 (34%) | [25][26][28][32][33][35][38][39][43][45][47][49][50][53][56][58][62][63][64][66][67][68][70][71] |
| Middleware | 20 (29%) | [27][29][34][37][39][41][44][47][48][50][53][56][57][59][61][63][65][68][69][71] |
| Modelling language | 12 (17%) | [27][28][33][37][41][44][45][46][54][59][60][61] |
| Comparisons | 6 (8%) | [35][51][52][57][60][72] |
| Tools | 4 (6%) | [36][40][46][62] |
| Ontologies | 4 (6%) | [30][31][42][55] |

Table 9 Primary study distribution according to the study's context

The study distribution indicates that methods, middleware and modelling languages are the contexts in which the choreography has been most frequently studied in the ES domain. A large number of studies under the methods category indicate that various development phases and artefacts are realised without or with limited methodological support. Recognising choreography in ES as a developing topic can be seen as one reason for this state and for the large number of studies in this category. One potential reason for large number of studies under the middleware and modelling language categories is a strong trend in the adoption of model-driven engineering (MDE) as an overall engineering approach to ES development [84]. In MDE, specific system viewpoints that are relevant to developers are specified using dedicated modelling languages, which are then translated to another specification or executed on middleware platforms. Understanding which concepts, rules and constraints are relevant to capturing the choreography viewpoint using modelling language and how to deploy and execute the specified choreography on a target middleware platform therefore represents an attractive research topic.

**Study motivation:** Six categories were derived to characterise the motivation for conducting the studies on choreography in the ES domain.

- Coordination gaps: Studies motivated by gaps in service coordination (interaction) mechanisms. Some examples of these gaps include the lack of a formal basis for coordination or the lack of support for the verification of coordination.
- Enterprise integration: Includes studies driven by the need to integrate low-level ES functionalities with high-level business functionalities into a comprehensive enterprise system.
- Lack of choreography language primitives: Includes studies motivated by the need to improve the choreography modelling languages in a way in which they are more applicable in the ES domain.
- Re-configurability: Studies motivated by the need to enable the rapid reconfigurations of ES.

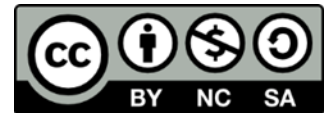

- Comparison: Studies in which two or more alternatives are compared to understand which alternative is more suitable in a given context. Some examples include the comparison of modelling languages, interactions schemes and techniques.
- Heterogeneity: Includes studies motivated by the problems caused by the use of heterogeneous hardware, operative systems, middleware platforms, modelling languages, communication protocols and standards.

The presented categories were derived based on the analysis of the problems addressed in the studies and represent a generalisation of those problems. Table 10 presents the distribution of the studies according to these categories.

| Motivation | Frequency (N=48) | Primary studies |
|---|---|---|
| Coordination gaps | 13 (27%) | [27][28][29][34][41][44][47][56][62][63][65][66][70] |
| Enterprise integration | 10 (21%) | [25][26][32][36][39][40][42][50][67][69] |
| Lack of language primitives | 7 (15%) | [37][38][45][46][55][59][61] |
| Heterogeneity | 7 (15%) | [48][49][54][58][71][64][68] |
| Comparisons | 6 (12%) | [35][51][52][57][60][72] |
| Re-configurability | 5 (10%) | [30][31][33][43][53] |

Table 10 Primary study distribution according to the motivation

Table 10 reveals that in the current literature, the coordination gaps, enterprise integration and the lack of language primitives are the most frequent problem areas addressed by choreography. One possible explanation for the number of studies in the coordination gap and the lack of language primitives categories is that ES domain imposes additional requirements on service coordination that are not addressed with the current coordination mechanisms and modelling languages. One potential explanation for the number of studies in the enterprise integration category is that enterprise-wide integration implies the cooperation of different systems used within the organisation. Choreography, as a viewpoint that captures the interactions of autonomous system parts, was seen as a potential tool that can facilitate the integration and an interesting topic for studying.

**Tools for choreography in the ES domain.** Two categories, languages and software packages, were derived based on the analysis of the studies to characterise the tool support for choreography in ES.

- Languages: present textual and visual languages that are used for specifying choreography in ES.
- Software packages: reveal software packages such as parsers, translators or editors that support the work with choreography languages.

The choreography tools data is organised in the matrix presented in Table 11. The identified languages are listed in the first row, while the software packages are listed in the first column. The studies reporting on the particular language and tool are marked in the intersecting cell of the matrix. Note that not all studies reported on the tools they used, that in several studies more than one language or software package was reported and that multiple occurrences of the study in a row or column were treated as a single study in frequency counts.

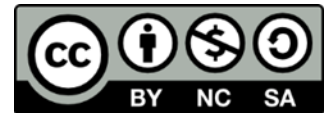

| Languages / Software packages | BPEL | UML, Java, .NET, Petri Net | BPMN | WS-CDL | SC-SCL | SALT | IEC 61* | PESM | SCRIBBLE | WSCI | ECL | APEL | CHOREO | BPEL4Chor | Frequency |
|---|---|---|---|---|---|---|---|---|---|---|---|---|---|---|---|
| Unspecified | [37] [39] [51] [58] | [33] [42] [64] [68] | [58] [71] | [59] [60] | | [48] [54] [69] | [60] [66] [67] | [33] [38] | [46] [56] | [35] | [27] | [41] | | | 22 |
| Eclipse | | | [47] [49] | [40] | [29] [45] [47] | | | | | | | | | [49] | 5 |
| Custom product | | | [55] | | | | | | | | | | [44] | | 2 |
| UML Model Transformation Tool | | [28] | | [28] | | | | | | [26] | | | | | 2 |
| ActiveWebflow Professional | [25] [32] | | | | | | | | | | | | | | 2 |
| Oracle JDeveloper | [25] | | | | | | | | | | | | | | 1 |
| Artisan Real Time Studio | | [28] | | [28] | | | | | | | | | | | 1 |
| **Frequency** | 7 | 6 | 5 | 4 | 3 | 3 | 3 | 2 | 2 | 2 | 1 | 1 | 1 | 1 | |

Table 11 Languages and software packages

The matrix revealed that the most commonly used software package is the Eclipse development environment (5 studies), but it also shows that the majority of the analysed studies did not report which software packages were used (22 studies). Eclipse is an open source product and freely accessible environment that offers a standardised way to develop various tools for tasks such as modelling, translating or verifying software lifecycle artefact [85]. These characteristics make Eclipse an attractive development environment that can be used both in industry and academia and represents a possible explanation for the large number of studies in this category.

BPEL [86], general purpose programing languages (for example UML, Java, .NET), business process model and notation (BPMN) [19] and WS-CDL [87] are the most commonly used languages for specifying choreographies in the ES domain. BPEL, BPMN and WS-CDL were used in 16 studies and together with WSCI and BPEL4Chor that are used in three studies represent the languages for specifying service interaction in the enterprise domain. UML, Java, .NET and Petri net were used in six studies and represent a well-known general purpose modelling language that can be used to capture service choreography. Strong commitment by researchers and practitioners to reuse the existing languages and knowledge in the ES domain is one potential explanation for the large numbers of studies in these categories.

SC-SCL, SALT, PESM, SCRIBBLE, ECL, APEL and CHOREO were used in 13 studies and represent custom developed or non-standardised choreography languages whose syntax and semantic is defined for the ES domain or it is tightly related to the ES development aspect they aim to support. There are two possible explanations. First, generic languages and choreography languages from the enterprise domain are not sufficiently expressive to capture the specifics of the ES domain. The second suggests that supporting some of the ES development aspects with dedicated choreography language constructs can result in various benefits, such as better comprehension, simpler development or verifiability.

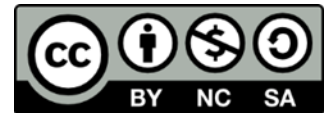

**Study impact.** The categories for characterising the impact of the results include the following:

- ES development process: reveals the impact on the ES development process.
- Quality: reveals the impact on the quality characteristics of the ES.
- Performance: reveals the impact on ES performance.

These categories were further refined and divided into subcategories that provide more focused characterisation of the impact and by differentiating which of the studies used quantitative approaches to verify the impact. Note that a single study can impact more than one category and the references of the studies whose impact is verified using a quantitative approach are marked with an asterisk sign. The distribution of the studies according to the derived categories is presented in Table 12.

| Category | Characteristic in focus | Frequency (N=69) | Primary studies |
|---|---|---|---|
| ES development process (29 studies) | Simplification of the ES development process | 18 | [27][28][29][31][32]* [38][40][41] [48][53][54][55][57]* [59][62][64] [68]* [69] |
| | Development speed | 6 | [25]* [26][40][48][54][69] |
| | Product integration aspects | 5 | [36][39][60][67][72] |
| Quality (24 studies) | Maintainability | 9 | [30][31][39][41][42][43][50][53][67] |
| | Verification & validation | 6 | [33][45][46] [47][56]* [61] |
| | Scalability | 5 | [41][44]* [49] [56]* [58] |
| | Reliability | 4 | [35]* [45][52]* [65]* |
| Performance (16 studies) | Shorter response time | 6 | [34]* [44]* [51]* [63]* [66]* [70]* |
| | Reduced power usage | 5 | [34]* [37][52]* [71]* [63]* |
| | Reduced bandwidth | 4 | [34]* [37][51]* [52]* |
| | Shorter network path | 1 | [52]* |

Table 12 Impact of choreography use in ES development

The study distribution reveals similar numbers of studies impacting the ES development process (24 studies) and quality aspects (29 studies), while performance is impacted in four studies. The simplification of the ES development process category contains a large number of studies (18 studies) since it was defined as a broader category, meaning that it includes contributions such as methods, middleware products and tools that can simplify the development process. Maintainability is the most commonly impacted quality characteristic (9 studies). One potential explanation for this is that in the context of service-oriented architecture, service interactions represent an essential means for handling various change requests, and thus the maintainability. Since choreography represents a viewpoint on service interactions, the impact of the results often targets this particular quality characteristic. Improvements in network utilisation and power consumption are reported in 16 studies and represent the two most common performance improvements for choreography use. One potential explanation for this is that choreography imposes a peer-to-peer interaction style between services. The peer-to-peer style omits the centralised coordination mechanisms, which is typical for orchestration, and enables direct interactions (choreography), thus impacting the response time and bandwidth.

Table 12 differentiates studies that used quantitative approaches to verify the impact of their results. The table illustrates that quantitative approaches are used more in studies whose results impact the performance category and only in several studies whose results impact the quality and ES development process. One potential explanation for this difference is the practical difficulty of measuring the impact on the quality and ES development process. This is particularly emphasised in cases in which the validation is conducted in an academic environment, which is the case for 67% of the studies, as presented in Appendix I.

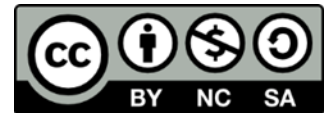

**Future trends:** Three categories indicating future research directions in ES choreography were derived from the studies.

- General improvements: Include studies that will continue to work on their artefact by fine-tuning, extending or by merging it with other development artefacts.
- Validation: Includes studies that reported activities such as experiments, measurement or pilots in industry as future work.
- Automation: Includes studies that will continue their research towards achieving the automatic selection and invocation of services, automatic reconfiguration and the automatic code generation based on their development artefact.

The study distribution according to future trends is presented in Table 13. Note that not all of the analysed studies reported their future work.

| Future trend | Frequency (N=37) | Primary studies |
|---|---|---|
| General improvements | 19 (51%) | [25][28][32][34][36][41][44][46][48][54][55][56][59][60][64][65][67][69][71] |
| Validation | 11 (30%) | [33][40][44][45] [49][50][52][54][57][59][61] |
| Automation | 7 (19%) | [31][39][48][49][56][66][70] |

Table 13 Primary study distribution according to future trends

As shown above, the majority of the studies that reported on future work will be directed towards general improvements of the developed artefact. Validation of the artefact is reported in 11 studies, which together with 19 studies from the general improvements category promise more empirical studies and reliable results of the impact their artefact has on ES development. Automation, as a future direction, is reported in 7 studies. Since it plays a significant role in model-driven engineering, which is increasingly present in ES development, it can be expected that the number of studies on automation will grow.

## 4.2   Choreography Adaption in the ES Domain

The results of choreography adaption in the ES domain were based on the analysis of 21 studies. These studies were selected since they provided sufficient data for the analysis of choreography adaption using the analytical tool. The analysis results consist of two parts. The first part presents the identified studies that discuss choreography adaption strategies and aspects, their distribution according to the analytical tool and their interpretation. The second part presents the comparison results of frequencies of the identified strategies and aspects in this study and in the Leite et al. [23] study. These results provide an answer to RQ2.

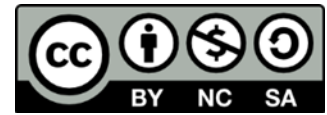

### 4.2.1 Choreography Adaption Strategies and Aspects in the ES Domain

Table 14 presents the studies organised according to the analytical tool and their frequencies, while the interpretations of these frequencies are presented in the following text.

| Aspects \ Strategies (N=29) | Targets (N=29) | | | Intervention degree (N=29) | | | Scalability (N=29) (3, 10%) | Implementation (N=29) (20, 69%) | Models (N=29) (20, 69%) |
|---|---|---|---|---|---|---|---|---|---|
| | Functional requirements (15, 53%) | Both (9, 30 %) | Non-functional requirements (5, 17 %) | Manual (12, 41%) | Hybrid (14, 49%) | Automated (3, 10%) | | | |
| Model-driven engineering-based (15, 53%) | [25][26][28] [33][38][40] [47][49][55] [67][71] | [36][64] [66][70] | | [25][36] [55][40] [64][66] [70][67] | [26][28] [33][38] [47][49] [71] | | [49] | [25][28][36] [40][47][55] [64][67] | [25][26] [28][33] [38][40] [47][49] [64][66] [67][71] |
| Middleware-based (7, 24%) | [37][48] [65][69] | [44][56] | [27] | [48] [65][69] | [37][44] [56] | [27] | [44][56] | [27][37][44] [48][56] [65][69] | [27][37] [44][48] [56][69] [67][68] [69] |
| Semantic reasoning-based (3, 10%) | | | [30][31][42] | | [30] | [31][42] | | [30] | |
| Measurement-based (2, 7%) | | [34] | [57] | [57] | [34] | | | [34][57] | |
| Multi-agent-based (1, 3%) | | [68] | | | [68] | | | [68] | [68] |
| Formal methods-based (1, 3%) | | [45] | | | [45] | | | [45] | [45] |

**Table 14 Choreography adoption strategies and aspects in the ES domain (adopted from [23])**

Table 14 reveals that the model-driven engineering based (15, 53%) and the middleware-based strategies (7, 24%) are the dominant choreography adaption strategies in the ES domain. These frequencies can be interpreted alongside the results from the analysis of the context, presented in Section 4.1.2, which showed that choreography was frequently studied in the context of middleware (20, 29%) and modelling languages (12, 17%). Since the models, their corresponding modelling languages and middleware products represent the cornerstones of the model-driven engineering approach, a similar explanation of the frequencies in this analysis can be proposed. This suggests that model-driven engineering, as an overall engineering approach, is increasingly present in the ES development in which choreography is utilised.

Table 14 also shows that the adaption strategies mostly target functional requirements (15, 53%). One potential reason for the large number of studies on functional requirements is that the choreography viewpoint aims to capture the interactions needed to achieve the system-level functionalities or goals that are the functional requirements of a system. However, addressing non-functional requirements with choreography can be seen as an important aspect since ES are often constrained with strict non-functional requirements, such as reliability, scalability, maintainability, real-time processing and performance.

The need to address non-functional requirements can also be augmented with the number of studies in which the choreography adaption is conducted in a hybrid manner (14, 49%), which, together with the automated intervention aspect (3, 10%), can be related to ES maintainability requirements. If the frequencies of studies in the hybrid adaption aspect are interpreted alongside the results from the impact analysis that is presented in Section 4.1.2 that revealed that the maintainability is the most impacted quality characteristic of a system, a similar explanation of the frequencies in this analysis can be proposed. This explanation suggests that hybrid, together with automated adaptations are frequently addressed in the studies since they are recognised as a means to improve the choreography adaption to various changes in the environment, and thus to improve the maintainability of ES.

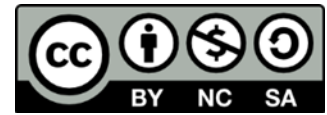

A significant number of the studies proposed some type of implementations (20, 69%) and used various models (20, 69%) in their research activities. These frequencies are aligned and can be explained with our observation regarding the presentation of research methods in the analysed studies. This observation was made during the quality assessment and suggests that the studies are design-oriented and strongly focused on solving practical problems using various artefacts. Examples of these artefacts include the implementation of tools, translators, languages and the utilisation of models and their representations.

### 4.2.2 Comparison of study frequencies

Leite et al. [23] characterised the choreography adaption strategies without considering the domain of application and for this purpose defined the analytical tool. This tool was used in this study to characterise the adoption strategies in the ES domain. The results from the comparison of the study frequencies in Leite et al.'s study and the frequencies from this analysis are presented in Table 15. The interpretation based on the comparison of each adaption strategy aspect is presented in the following text.

| Adaption strategy aspect | | Leite et al. (N =24) | This study (N=29) | Difference | |
|---|---|---|---|---|---|
| | | | | Count | % |
| Target | Functional | 11 (46%) | 15 (53%) | < 4 | < 7% |
| | Non-functional | 2 (8%) | 5 (17%) | < 3 | < 9% |
| | Both | 11 (46%) | 9 (30%) | > 2 | > 16% |
| Human intervention degree | Automated | 16 (67%) | 3 (10%) | > 13 | > 57% |
| | Manual | 4 (17%) | 12 (41%) | < 8 | < 24% |
| | Hybrid | 4 (17%) | 14 (49%) | < 10 | < 34% |
| Scalability | Experiments | 1 (4%) | 2 (7%) | < 1 | < 3% |
| | Informal | 3 (12%) | 1 (3%) | > 2 | > 9% |
| | Not reported | 20 (83%) | 26 (90%) | < 6 | < 7% |
| Implementation | Yes | 14 (58%) | 20 (69%) | < 6 | < 11% |
| | No | 10 (42%) | 9 (31%) | > 1 | > 11% |
| Model | BPMN | 1 (4%) | 3 (11%) | < 2 | < 7% |
| | WS-CDL like | 2 (8%) | 3 (11%) | < 1 | < 3% |
| | UML | 3 (12%) | 1 (3%) | > 2 | > 9% |
| | Process algebra | 2 (8%) | 1 (3%) | > 1 | > 5% |
| | Automata | 2 (8%) | 0 (0%) | > 2 | > 8% |
| | Other models | 3 (12%) | 12 (41%) | < 9 | < 29% |
| | Not reported | 10 (48%) | 9 (31%) | > 1 | > 17% |

Table 15 Adaption strategy analysis results

The frequency comparison of the target indicates that in the ES domain, more effort is made to investigate the adaption of non-functional requirements. Compared with Leite et al.'s [23] findings, the number of studies that investigate the non-functional requirements are 16% higher in this study. The stronger emphasis on non-functional requirements can be explained by the specific constraints that exist in the ES development domain and impose stronger consideration of non-functional requirements. Examples of these constraints are discussed in [6], [13] and include limitations in processing power, storage and power supply, the demands for reliability, scalability, maintainability, real-time processing and execution in ad-hoc network environments.

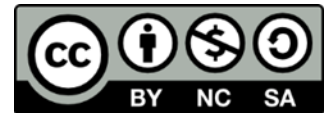

A human intervention degree analysis reveals that the strategies with the hybrid intervention degree are the dominant strategies in the ES domain (14, 49%), which differs from Leite et al.'s [23] study in which automated intervention strategies is dominant (16, 67%). Lower utilisation of automated techniques in ES can be explained by the relatively short (10-year) time span of choreography usage in the ES domain. During this time, the automation techniques used for choreography adaption could not reach higher degrees of maturity. The analysis of the future trends presented in Section 4.1.2 shows that 7 studies plan to continue their research by automating their current proposals. This indicates that more studies focusing on automation can be expected in near future.

Scalability is the aspect of choreography adaption that has not yet been studied sufficiently in the ES domain; it is only considered in 3 studies, or 10% of all analysed studies. These results are similar to those of Leite et al. [23] in which 4 studies, or 16%, discuss scalability. Accordingly, a conclusion similar to that of Leite et al. [23] can be drawn, suggesting that exploring the impact of the choreography adaption strategy on scalability represents an attractive research topic. This is particularly interesting in the ES domain due to rapid increase of their size.

In terms of implementation, a significant number of the studies (58% and 69%) analysed in both SLRs include some type of implementation, such as tools, translators or code generators. Based on these frequencies, the same explanation proposed in this study can be proposed for studies analysed by Leite et al. [23]. This suggests that the studies are design-oriented and strongly focused on solving practical problems using various implementations.

The analysis of the underlying models and languages revealed that standardised models, such as BPMN, WS-CDL like, UML, process algebra and automata, are similarly used in studies analysed in both SLRs (40% and 28%) and that the utilisation of other, custom-made or non-standardised, models is represented more in studies analysed in this SLR (12% and 41%). Standardised models, such as BPMN and WS-CDL, are designed with enterprise system development in mind. One potential explanation for similar percentages of standardised models in both SLRs is that the majority of the studies in Leite et al. [23] focus on enterprise systems development, while in the ES domain a number of studies addressed their problems by transferring and using models from the enterprise domain. In both cases, the selection of the standardised languages represents a logical choice. Custom or non-standard models are more often represented in this SLR. One potential explanation for this is partially aligned with the problem that motivated this review, which is the insufficient expressiveness of existing models to capture the complexities that are typical in the ES domain. Accordingly, new languages or customisation of existing languages are being developed to address this problem in the ES domain.

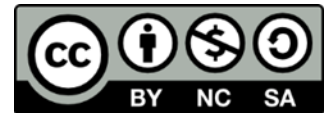

## 4.3   Utilisation Types of Choreography Specification in ES Development

A choreography specification can be used in different ways and in different stages of the ES development process. The analysis of the studies revealed three types of choreography specifications used: descriptive, parsed and executable specification.

- *Descriptive* represents a conventional or traditional understanding of specifications in which choreography is used as an analytical tool for reasoning and documenting the global service interaction scenarios. This type of specification is developed in the early phases of ES development using informal or semi-formal methods. For the development and maintenance of this specification type, different modelling languages and specialised tools can be used, but this is not a mandatory requirement. Conventional tools, such as text processors, slides and free-form descriptions, are also a viable alternative.
- *Parsed* is a type of specification that can be read or parsed by the development tools. Compared with the descriptive specification type, reading or parsing of a specification extends its use by enabling various features, such as code generation. Parsed choreography specifications are used during the ES design phases and use the formal modelling languages and specialised tools, which are often part of the larger tool chain to describe the service interactions.
- *Executable* is a type of specification that controls the behaviour of a running system. Compared with the parsed specification, which is parsed by the development tools, executable specifications are executed by the specialised middleware component during the system runtime. Accordingly, the development of these specifications requires formal language, tool support and the middleware functionality capable of executing such specification.

The distribution of the studies according to specification types is presented in Table 16, which is followed by the short summary of a representative study from each type and the interpretation of the distribution.

| Specification type | Frequency (N=33) | Primary study |
|---|---|---|
| Parsed | 21 (64%) | [28][33][34][35][38][40][41][44][45][47][49][51][54][56][57][60][64][65][67][69][71] |
| Executable | 7 (21%) | [27][37][42][58][66][68][70] |
| Descriptive | 5 (15%) | [25][26][32][61][55] |

**Table 16 Choreography specification types**

The most frequently used specification type is the parsed specification type, reported in 21 studies. A *parsed* specification includes, among others, the study by Cambronero et al. [28], which introduces the development method-based UML profile for real-time systems modelling (RT-UML). One of the features of this method is the transformation of RT-UML models into WS-CDL language that includes details about time constraints, which is then translated into BPEL. The *executable* specification type is used in seven studies. An example study in which this type of specification was used is the work of Zhang et al. [37]. They extended the BPEL syntax with the attributes containing choreography information. Once the execution is initiated, the composition specification is forwarded through the execution path informing each service provider about the composition. The least frequently used specification type is the *descriptive* choreography specification type, used in five studies. For example, March et al. [61] proposed a workflow model consisting of abstractions specific for mobile applications that interact with cloud services. In this context, a choreography specification describes the sequence of interactions between the applications running on mobile devices and clouds.

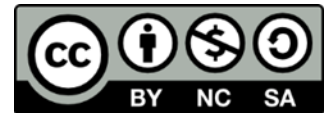

Reading or parsing a specified choreography model can result in several benefits during ES development, which were seen as possible explanations for the number of studies in the *parsed* category. The benefits from a specification that is readable by tool include: (a) automatic generation of the subsequent specification or specification parts, for example in the model-driven engineering context, (b) code generation (c) traceability between constructs in different specifications, (d) data extraction for the purposes of the specification improvements and (e) verification of various specification properties such as dead-lock freeness.

## 4.4   Realisation Types of Choreography Specification in the ES Domain

To realise the specification refers to how the specified choreography is conducted (or enacted) during the system execution. Three types of choreography realisation were identified during the analysis of the studies: participant driven, distributed and piggybacked.

- *Participant-driven*: The specified choreography is divided into fragments by specialised tools during the design phase and deployed to corresponding participants. During the ES execution, choreography is not executed per se, but realised as a consequence of the parallel execution of participants' application fragments and their message exchange. In this case, fragments of information from the choreography specification are built in the participant's logic.
- *Deployed:* The specified choreography is deployed to a central choreography engine or distributed to all participants in the system. In this way, a choreography specification is executed centrally or locally in the execution environment of each participant.
- *Piggybacked:* The specified choreography is carried or piggybacked from one participant to another as part of the messages that are being exchanged. In this way each participant has up-to-date information about the choreography and realises the message exchange accordingly.

The distribution of the studies according to realisation types is presented in Table 17, which is followed by the short summary of a representative study from each type and the interpretation of the distribution.

| Organisation types | Frequency (N=28) | Primary study |
|---|---|---|
| Participant driven | 20 (71%) | [28][32][33][34][38][41][44][45][47][51][54][56][60][61][64][65][66][67][69][70] |
| Deployed | 6 (21%) | [42][49][57][58][68][71] |
| Piggybacked | 2 (8%) | [27][37] |

**Table 17 Choreography organisation types**

*Participant-driven* realisation of the choreography specification is the most frequently used type of realisation, reported in 20 studies. An example study in which this realisation type was used is the work of Sen et al. [34] that proposes the choreography-based workflow system designed for mobile networks and devices. This system divides the global workflow into fragments using its own allocation algorithm, which relies on available information about the available devices. Once the fragmentation is done, the workflow system deploys the fragment to the devices for execution. A *deployed* realisation of a choreography specification is identified in six studies. For example, Herrera et al. [58] proposed a model for collaborative mobile services for the Internet of things. The model assumes that all devices engaged in collaboration acquire knowledge about global interactions by sharing the global workflow, which is the choreography. *Piggybacked* realisation is the least frequently used realisation type, and it is reported in two studies. One example is the study by Cottenier and Elrad [27] that proposed a comprehensive framework for embedded services. In this framework, the choreography information is included in the header of the messages that are exchanged between participants.

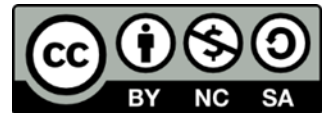

The *participant-driven* type, indicating that choreography is realised as a consequence of the parallel execution of participants' services, is the dominant realisation type in the studies. One potential explanation for this is that the *participant-driven* type follows the general or conventional understanding of choreography according to which choreography shows the service interactions from the global point of view that progresses as individual participants are exchanging messages. As a proxy to the conventional understanding of choreography, this type of realisation was identified in most of the analysed studies.

## 4.5   Implicit Assumptions for Choreography Use in the ES Domain

Two implicit assumptions were identified during the analysis as important for the utilisation of choreography in the ES development domain. These implicit assumptions are participant autonomy and network stability. The main reason these two implicit assumptions were seen as important is the identified differences in how they are interpreted in ES compared with the non-ES domain. These differences and the studies in which they are identified are presented in this section.

**Participant autonomy** denotes that participants engaged in the choreography scenario represent distinct administrative authorities. Examples of participants in the enterprise domain include organisations, companies or autonomous organisation units within large corporations. In the ES domain, participant examples include a device in the production line, a group of sensors and actuators attached to a single controller or an electronic controller unit (ECU) in an automobile. Following the participant autonomy assumption, each participant implements and executes its own functionalities (for example, as orchestrations) while the message exchange between them advances according to the specified choreography. The analysis of the studies, however, revealed that in the ES domain this is not always the case. In the ES domain, participants' services engaged in choreography are often administered by a single authority. In this way it becomes challenging to differentiate choreography and orchestration as global and local viewpoints to interactions since there is a single authority and therefore a single participant. Instead of complementing global and local viewpoints, choreography and orchestration are interpreted as competing viewpoints that denote decentralised and centralised ways of organising the interactions.

To understand the underlying reasons for the interpretation differences, the studies were classified according to whether or not they are aligned with the participant autonomy assumption. These studies are presented in Table 18.

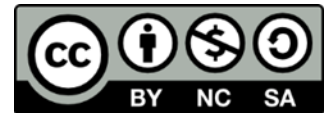

| Participant autonomy | Frequency (N=36) | Primary studies |
|---|---|---|
| Not aligned (single authority) | 20 (55%) | [33][34][38][39][41][44][46][48][51][53][56][60][61][62][65][66][67][68][69][70] |
| Aligned (multiple authorities) | 16 (45%) | [25][26][27][28][31][32][37][42][45][47][49][57][58][63][64][71] |

**Table 18 Participant autonomy**

One possible reason ES developers use choreography in cases in which all participants are administrated by a single authority is the need to increase or retain the overall system performance, which is important due to various resource limitations. Accordingly, to instead use the orchestration engine and implement a client-server interaction scheme between the participants, ES developers design the system following the choreography scheme, which implements peer-to-peer interactions between participants. By implementing a peer-to-peer interaction scheme, ES developers omit the use of orchestration engines and gain performance improvements, such as shorter network paths, reduced power consumption [52] and increased execution speed [51].

The drawback of implementing choreographed (peer-to-peer) interactions is the decrease in maintainability since the interaction logic has to be deployed to each participant involved in the scenario. For example, when a change request is introduced, the ES developer must identify all affected participants and update their code. This can take more time and effort compared to cases in which the interaction logic is maintained and executed centrally, using orchestration engines. In this case, an ES developer modifies the orchestration script according to the change request and deploys it to the orchestration engine. This drawback however, is recognised in ES development communities; as a response, development environments and middleware products that support the central management of choreography have been proposed. These central management systems for choreography are reported in [27][44][53][56] and [62].

**Network stability** denotes that participants engaged in choreography scenarios are known and defined during the system's design. A network formed by interacting participants is therefore seen as stable if participants are not expected to disconnect from the network, nor are additional participants expected to join the network during the system's execution. The study analysis however, revealed that in the ES domain the network of participants engaged in choreography is not always stable and can change during the system's execution. In these cases, participants can freely join or leave the network, forming what is known as an ad-hoc network.

To understand the underlying reasons for the identified differences, the studies were classified according to whether or not they are aligned with the network stability assumption. These studies are presented in Table 19.

| Network stability | Frequency (N=36) | Primary studies |
|---|---|---|
| Stable network | 29 (80%) | [25][26][27][28][32][33][38][39][41][45][46][47][48][49][51][53][56][57][60][61][62][64][65][66][67][68][69][70][71] |
| Ad-hoc network | 7 (20%) | [31][34][37][42][44][58] [63] |

**Table 19 Network stability**

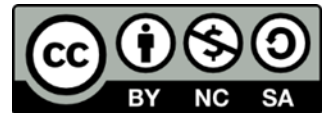

The analysis of the studies listed in Table 19 revealed that ad-hoc networks are formed in telecommunications and sensors & actuators industry branches. In the telecommunications branch, a participant is commonly identified by a mobile device that is wirelessly connected to the network. The mobility of the device enables participants to freely move in and out of the area covered by the network and allows the possibility for joining or leaving the choreography scenario that is being executed. The enablement of ad-hoc networking in telecommunications can therefore be seen as a consequence of the mobility or motion of a device and therefore needs to be considered during the specification of the choreography scenario.

In the sensors & actuators branch, a participant is represented with a dedicated device that controls one or more sensors and actuators, and multiple participants are connected via wired or wireless networks. Systems that rely on sensors and actuators are often executed in environments that impose high reliability requirements, such as factories, aeroplanes, automobiles or smart buildings. The execution of these systems therefore must not be interrupted even if some of the participants get disconnected from the network due to, for example, device failures. In addition, the joining of new participants to the network, for example, due to the expansion of the network coverage, must not interrupt the execution of the system. In the sensors & actuator branch, participants therefore express the characteristics similar to mobile devices due to the properties of the environment where they are executed which leads to the emergence of ad-hoc networking.

# 5  Study Findings

Two groups of findings were derived from the results. The first includes the findings oriented towards research and reveals the knowledge gaps that can be seen as interesting for future research. These topics discuss choreography as a pervasive specification, languages for choreography modelling, adaption strategies, and empirical studies and synthesis. The second group includes the findings oriented towards practice and reveals areas that can be useful to experts who utilise choreography in ES development. These topics discuss the participants in ES development, performance and technological heterogeneity. These findings are seen as current features of the choreography use in ES domain.

## 5.1  Research-Oriented Findings

**Choreography as a pervasive specification.** The analysis revealed three types of choreography specifications and that each type is used at different ES development stages — analysis, design and execution. Instead of being tied to a single stage, the results indicate that choreography is pervasive throughout the ES development process and consequently provides an opportunity to study the needed requirements and the effects they produce in different development stages. In addition, the relationship of a choreography specification with other development activities and artefacts can also be seen as an interesting research area, particularly in the model-driven engineering context, which is increasingly present in ES development.

**Languages for choreography modelling.** The analysis of the study's motivation revealed that choreography modelling languages are still not sufficiently expressive to capture the complexities that are typical in ES domain. This research opportunity can be addressed by following one of the two identified streams. The first stream is to use an existing language for choreography modelling and customise it according to the ES domain needs, while the second is to build a language from scratch. Previous studies on this topic from both streams are presented in Section 4.1.2 and seen as a valuable source of information for researchers engaged in language development.

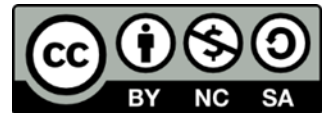

**Adaption strategies.** The analysis of choreography adaption, as a property of the specification that enables it adjustments to changes, revealed that the multi-agent-based strategy, scalability and automation are insufficiently explored. The multi-agent-based strategy enables ES to learn from previous tasks and to autonomously act based on what was learned. Scalability, the ability of ES to perform during increases in workload, and automation, the ability to adapt to changes without human intervention, are seen as favourable aspects of each adaption strategy since the size and complexity of ES is constantly increasing. Thus, all three seem to be interesting areas for future research.

**Empirical studies and synthesis.** The quality assessment of the studies revealed that a limited number of studies explored the application of choreography in practice and that the majority of studies were reported without considering the synthesis of their results in future systematic literature reviews. More empirical studies in the industry context with consideration of synthesis in their reports facilitates later derivations of evidences, thus are highly emphasised. To this end, Wohlin [88], who identified terminology, content and review of empirical studies as areas relevant for knowledge synthesis in future SLRs, made recommendations that represent a good starting point.

## 5.2 Practice-Oriented Findings

**Participants in ES development**. The study analysis revealed differences in how participants are understood in ES and non-ES domains. In the ES domain, a choreography specification often captures service interactions that are administered by a single management authority, also known as a participant. This is different than enterprise systems in which the choreography captures service interactions among participants and where each participant is administrated by a distinct management authority. Instead of understanding the choreography and orchestration as two complementing viewpoints that capture the interactions from two different abstraction levels, in the ES domain two representations often differ based on whether they capture the client-server (orchestration) or peer-to-peer (choreography) type of interactions. In this way, choreography and orchestration do not complement each another, but represent two competitive approaches to how service interactions can be organised. Understanding whether participants are under single or multiple management authorities is seen as highly important in ES development since it determines how choreography can be used, and thus how it contributes to ES development.

**Performance improvements.** The study analysis indicates that ES can gain performance improvements, such as shorter network paths, reduced power consumption and a faster system response, if the interactions in their ES are organised according to choreography. The ability to achieve these improvements, however, strongly depends on the implicit assumption regarding participant autonomy. For example, if all the participants are managed by a single authority, their interactions can be organised according to choreography, which improves the ES performance. Additionally, the ownership of the participants allows the use of a common middleware product, thus enabling easier development due to the uniform platform and the possibility of executing the choreography scenarios.

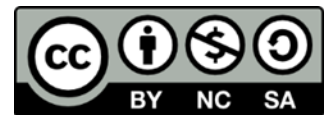

**Technological heterogeneity.** Technological heterogeneity is related to participants that are built using different technologies (hardware, network protocols and operative systems) and implies that multiple management authorities are administrating the participants involved in the interactions. In cases in which the interacting participants use heterogeneous technologies, a choreography specification can be used during the analysis and design to capture the technological data about participants involved in the scenario. In this way, differences in the implementation technology become visible earlier in the development, thus providing upfront knowledge for developers about constraints imposed by those technologies. Heterogeneous technologies and multiple management authorities often omit the possibility to execute the choreography specification, unless all management authorities agree on a common execution environment. If the common execution environment cannot be used, choreography can only be used as a descriptive or parsed specification type.

# 6 Threats to validity

The threats to validity that were addressed in this study include descriptive, theoretical and conclusion validity, generalisability and repeatability. These validity threats, according to Petersen and Gencel [89], are applicable in the software engineering field and adopted in this SLR following the recommendations by Petersen et al. [90], who applied them to ensure the validity of systematic mapping studies. Since systematic mapping studies and systematic literature reviews are based on a similar idea, which is to thoroughly explore the existing literature and to produce new knowledge based on the selected literature, addressing these validity threats was seen as reasonable for this study's purpose.

Descriptive validity is related to the extent to which observations are objectively described [89]. In systematic literature reviews, the threat to descriptive validity is related to the accuracy of the description of the data extracted from the primary studies [90]. This threat was addressed by planning, extracting and thoroughly describing the data relevant for the stated RQs and by using a specialised tool during these tasks. The extracted data were categorised into the bibliography, research, strategy and adoption data groups. NVivo, a specialised software package for qualitative analysis [24], was used to support the extraction, storage and management of the extracted data.

Theoretical validity is related to the appropriateness of the research approach to derive the intended outcomes [89]. In systematic literature reviews, a threat to theoretical validity is related to the extent to which the implementation of the search protocol is likely to result in studies that are relevant to the study's goal [90]. Theoretical validity was addressed by applying the following countermeasures: a) a thorough description of the rationale behind the keywords used for the search string is provided, b) the search string that was derived based on the keywords was piloted and the search results were checked against the studies know to researchers, c) two researchers were involved in deciding whether the publication is suitable for this study and d) reference lists of the selected primary studies were manually checked to identify additional relevant studies.

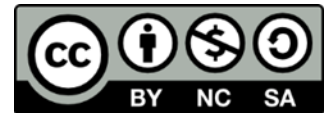

Conclusion or interpretation validity is related to the association between the results and the data based on which those results were derived [89]. In systematic literature reviews, threats to conclusion validity include subjective decisions that may occur during the selection of primary studies and data extraction [90]. Conclusion validity was addressed by applying the following countermeasures: a) all involved researchers were familiar with the search protocol, b) at least two researchers were involved in the selection of the studies based on reading the abstract and based on scanning the introduction, conclusion and illustrations, c) criteria used during the selection were piloted beforehand and adjusted based on the pilot results and d) all the disagreements over the selections were handled through the joint discussions.

Generalisability is related to the extent to which the study results can be generalised, or applied in different contexts [89]. In systematic literature reviews, generalisability is related to the extent to which the selected primary studies can represent the topic in focus [90]. To ensure generalisability, we applied the following countermeasures: a) a systematic literature review method that is well-accepted in the software engineering field was adopted and rigorously implemented, b) broad coverage of the literature sources was achieved by searching eight relevant databases, and c) the reference lists of the selected studies were manually checked to additionally support the literature search.

Repeatability is related to the ability to replicate a study process and to achieve identical results [89]. In systematic literature reviews, a threat to reliability is related to the possibility of obtaining different sets of primary studies if the study process is replicated [90]. To address this threat, a thorough description of the implementation of the search protocol is provided. This makes it possible to replicate this study with the same outcomes.

This study also runs the risk of positive publication bias. Positive publication bias suggests that studies with negative results are never or rarely published, making the search process and the selected studies biased towards positive results. Although Petersen et al. [90] considers publication bias to be part of theoretical validity, this threat was presented separately since it is inherent in studies of this kind and consequently not addressed in this review.

## 7   Conclusion

The objective of this study was to identify current features of the use of choreography in the embedded systems domain that are relevant for practitioners and researchers. For this purpose, a systematic literature review was conducted. During the review, 48 primary studies focusing on choreography use in the ES development domain were selected from the initial set of 6823 studies and analysed using the thematic synthesis approach. A thorough analysis of these studies resulted in a comprehensive characterisation including the classification of the studies, research reported in those studies, description of the choreography adaption strategies in the ES domain, identification of choreography specification and realisation types and the discussion of two implicit assumptions that should be considered if using choreography in the ES domain. The findings provide new knowledge that can be used to improve ES development.

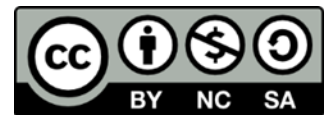

The classifications of the studies revealed that conference proceedings account for the most frequently used publication channel, industry automation and telecommunications represent two industry branches in which choreography is most frequently used and that the tool support relies mostly on Eclipse-based technologies and on modelling languages originating from the enterprise systems domain. Further characterisation of the studies indicates that choreography is seen as a solution for problems caused by technical and technological heterogeneity. Aligned with this is finding from the analysis of the study context, which revealed that middleware is the context in which choreography is most frequently studied. These two results can be seen as complementary since one of the main purposes of middleware is to hide the complexity caused by heterogeneous hardware, operative systems and network technology. Quality assessment of the studies has shown that the results are mainly validated in academic settings. In the future, however, more industry cases can be expected due to the international research projects that focus on the use of choreography in the ES domain, such as CHOReOS and CHOReVOLUTION.

The assessment of choreography adaption revealed that the most frequently used adaption strategy is the model-driven engineering-based strategy, while the agent-based strategy and scalability and automation as two strategy aspects are not sufficiently explored areas and more research can be allocated to them in the future. A comparative analysis between the study frequencies in Leite et al. [23], who studied the choreography adaption in the non-ES domain, and our study indicates that in the ES domain there is a stronger emphasis on non-functional requirements, a stronger tendency towards the use non-standardized modelling languages for specification and an equally frequent use of implementations as a mean to verify the proposals.

Four themes should be considered for the successful utilisation of choreography in ES. These themes emerged from the study analysis and include choreography specifications, assumptions, realisation and middleware functionality. The specification theme revealed three types of choreography specification use: descriptive, parsed and executable. Two choreography assumptions, participant autonomy and network stability, revealed that in the ES domain, participants do not always represent autonomous authorities and that participants often form ad-hoc networks in the ES domain. The realisation theme revealed that during execution, or system run-time, choreography can be participant driven, deployed and piggybacked.

The results derived in this study support the choreography use that potentially leads to various improvements in the ES domain. By explaining the identified characteristics and by revealing research opportunities and means for gaining practical improvements, the results facilitate the adoption of choreography among ES practitioners and enable further research advancements.

## Acknowledgements

This study is part of AMALTHEA and DIGILE – N4S (Need for Speed) projects. Both projects are funded by Tekes the Finnish Funding Agency for Innovation. The authors are also grateful to Professor June Verner for providing guidance and constructive feedback and to Qianhui Zhong for helping during the realisation of the search protocol.

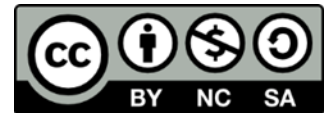

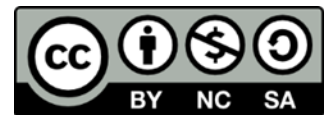

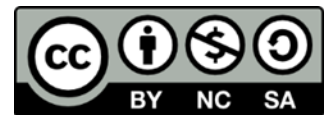

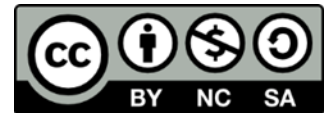

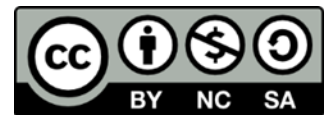

## Appendix I

Three criteria that were used in Nicolas and Toval SLR [21] were adopted for the quality assessment in this study. The first criterion, the publication sources, was satisfied by selecting databases that are known to publish the scientific literature in the software engineering field. This criterion was also additionally supported by the search protocol that ensured that all studies are peer-reviewed before publishing. The second criterion, or the artefact support, was satisfied by ensuring that the studies focus on choreography or on artefacts related to choreography in the ES domain. The compliance with this criterion was also supported by the search protocol, during which the researchers scanned the content of the publications and retained only those that address this topic. The third criterion, validation procedures, was satisfied by ensuring that the validation procedures used in the studies match one of the validation levels defined by this quality criterion. These validation levels are as follows:

- The study is validated in an academic environment or using examples
- The study is validated using known research methods
- The study has been validated as part of an industrial case study

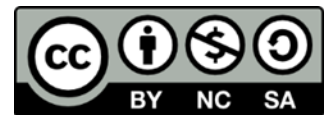

- The study can be seen as part of industrial practice

The distribution of studies according to these levels is presented in Table 20.

| Levels of validation procedures | Frequency (N=48) | Primary studies |
|---|---|---|
| Academic environment or example | 32 (67%) | [26][27][28][29][30][31][33][36][37][38][40][42][46][47][48][49][45][54][58][50][55][61][62][63][64][65][66][67][68][69][70][71] |
| Research method | 8 (17%) | [25][32][34][35][44][52][57][59] |
| Industrial case | 6 (12%) | [29][39][43][53][60][72] |
| Industrial practice | 2 (4%) | [41][56] |

**Table 20 Primary study distribution according to validation procedures**